\documentclass[twocolumn,prl,floatfix,superscriptaddress,nobibnotes]{revtex4-1}

\usepackage{amsmath}
\usepackage{amssymb}
\usepackage{graphicx}
\usepackage{color}

\begin{document}

\title{Anomalous Metallic Phase in Tunable Destructive Superconductors}

\author{S.~Vaitiek\.{e}nas}
\affiliation{Center for Quantum Devices and Microsoft Quantum Lab--Copenhagen, Niels Bohr Institute, University of Copenhagen, 2100 Copenhagen, Denmark}
\author{P.~Krogstrup}
\affiliation{Center for Quantum Devices and Microsoft Quantum Lab--Copenhagen, Niels Bohr Institute, University of Copenhagen, 2100 Copenhagen, Denmark}
\author{C.~M.~Marcus}
\affiliation{Center for Quantum Devices and Microsoft Quantum Lab--Copenhagen, Niels Bohr Institute, University of Copenhagen, 2100 Copenhagen, Denmark}

\date{\today}

\begin{abstract}
Multiply connected superconductors smaller than the coherence length show destructive superconductivity, characterized by reentrant quantum phase transitions driven by magnetic flux. We investigate the dependence of destructive superconductivity on flux, transverse magnetic field, temperature, and current in InAs nanowires with a surrounding epitaxial Al shell, finding excellent agreement with mean-field theory across multiple reentrant transitions. Near the crossover between destructive and nondestructive regimes, an anomalous metal phase is observed with temperature-independent resistance, controlled over two orders of magnitude by a millitesla-scale transverse magnetic field. 
\end{abstract}

\maketitle

Quantum phase transitions (QPTs) \cite{Sondhi1997,Vojta2000} in conventional superconductors serve as prototypes for related effects in more complex, strongly-correlated systems \cite{Shah2007}, including heavy-fermion materials \cite{Si2010} and high-temperature superconductors \cite{Norman2011}. While low-temperature superconductors are well understood in bulk, new phenomena can arise in mesoscopic samples and reduced dimensionality \cite{Tinkham1996,DelMaestro2009}. For instance, in two-dimensional films, electrons theoretically condense into either a superconductor or insulator in the low-temperature limit \cite{Goldman2010}. Yet, in many instances, an anomalous metallic state with temperature-independent resistance is found at low temperatures \cite{Kapitulnik2019}. In one-dimensional wires, incoherent phase slips can destroy superconductivity \cite{Bezryadin2000} or give rise to an anomalous metallic state \cite{Zaikin1997}, while coherent quantum phase slips can lead to superposition of quantum states enclosing different numbers of flux quanta \cite{Astafiev2012}, potentially useful as a qubit \cite{Mooij2005}.

Multiply connected superconductors provide an even richer platform for investigating phase transitions. Fluxoid quantization in units of $\Phi_0 = h/2e$ \cite{Deaver1961,Doll1961}, reveals not only electron pairing but a complex macroscopic order parameter, $\Delta e^{i\varphi}$ \cite{Tinkham1996,Douglass1963}. The same physical mechanism underlies the Little-Parks effect, a periodic modulation of the transition temperature, $T_{\rm C}$, of a superconducting cylinder with magnetic flux period $\Phi_0$ \cite{Little1962}.  For hollow superconducting cylinders with diameter, $d$, smaller than the coherence length, the modulation amplitude can exceed zero-field transition temperature, $T_{\rm C0}$, leading to a reentrant destruction of superconductivity near odd half-integer multiples of $\Phi_0$ \cite{deGennes1981,Schwiete2009,Arutyunyan1980}. 

Early experimental investigation of the destructive Little-Parks effect reported reentrant superconductivity interrupted by an anomalous-resistance phase around applied flux $\Phi_0/2$ \cite{Liu2001}. Subsequent experiments showed a low-temperature phase with normal-state resistance, $R_{\rm N}$, around $\Phi_0/2$, but did not display fully recovered superconductivity at higher flux \cite{Sternfeld2011}. Several theoretical models were proposed to interpret these  different scenarios \cite{Vafek2005, Lopatin2005, Dao2009}, but no consensus emerged.

\begin{figure}[b!]
\includegraphics[width=\linewidth]{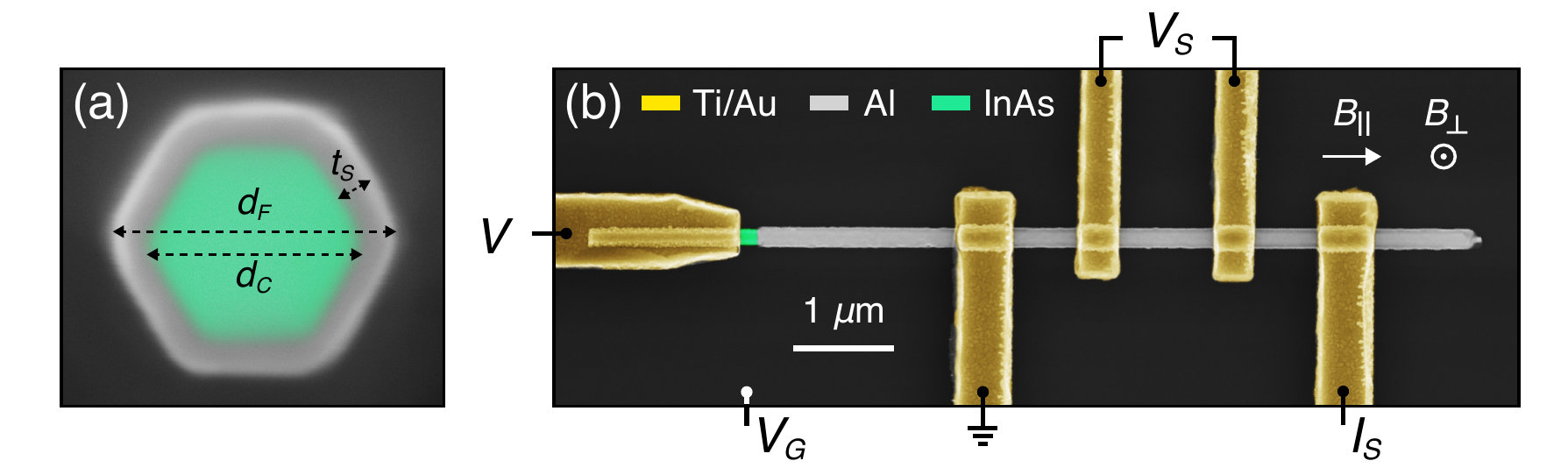}
\caption{\label{fig:1} (a) Colorized material-sensitive scanning electron micrograph of InAs-Al hybrid nanowire cross-section. The full wire diameter $d_{\rm F}$, core diameter $d_{\rm C}$ and shell thickness $t_{\rm S}$ are indicated by dashed arrows. (b) Representative color-enhanced micrograph of a device (wire B) consisting of an InAs core (green) with Al shell (grey), contacted with Ti/Au leads (yellow). The device can be operated in voltage ($V$) and current ($I_{\rm S}$) bias measurement set-ups.}
\end{figure}

Here, we report a study of the Little-Parks effect across the transition from destructive to nondestructive regimes, in InAs nanowires with a thin epitaxial cylindrical Al shell.  Remarkable agreement with Ginzburg-Landau mean field theory is observed across multiple reentrant lobes as a function of flux, temperature, and current bias, using independently measured material and device parameters. We then investigate a field-tunable crossover from non-destructive to destructive regime. At the boundary, an anomalous metal phase is identified, characterized by a temperature-independent resistance that can be tuned over two orders of magnitude using small changes in perpendicular magnetic field, $B_\perp$. We interpret these results in terms of tunneling between adjacent fluxoid states with different phase winding numbers giving rise to an anomalous metallic phase. As noted previously \cite{Vafek2005}, the appearance of a field-tunable temperature-independent resistance does not emerge naturally from simple models. The basic mechanism leading to a field-tunable saturating resistance remains mysterious. 

\begin{figure}[t!]
\includegraphics[width=\linewidth]{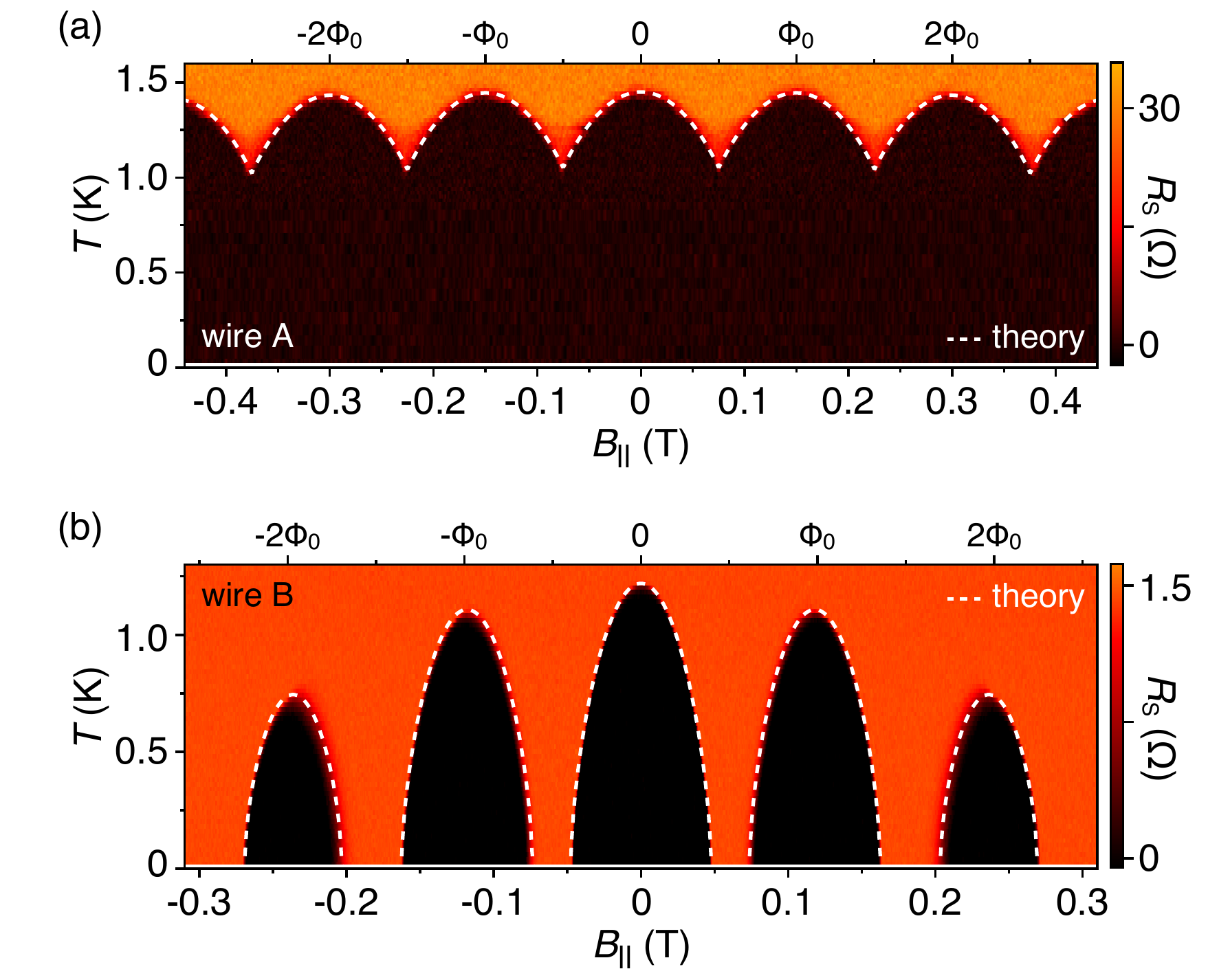}
\caption{\label{fig:2} (a) Shell resistance, $R_{\rm S}$, measured for wire A with shell thickness $t_{\rm S}=7$~nm as a function of axial magnetic field, $B_\parallel$, and temperature, $T$. The superconducting transition temperature of the shell is periodically modulated by $B_\parallel$. The sample is superconducting for temperatures below 1~K throughout the whole measured $B_\parallel$ range. The dashed theory curve is Eq.~\ref{eq:digamma} evaluated with $\alpha_\parallel$ from Eq.~\ref{eq:alpha_paral} and the corresponding fit parameters measured for the wire~A. (b) Same as (a), but measured for wire B with shell thickness around $t_{\rm S}=25$~nm, showing the destructive regimes around $\pm \Phi_0/2$ and $\pm 3\Phi_0/2$ of the applied flux quantum. }
\end{figure}

The devices we investigated were made using InAs nanowire grown by the vapor-liquid-solid (VLS) method using molecular beam epitaxy (MBE). Following wire growth, an epitaxial Al layer was grown within the MBE chamber while rotating the sample stage, resulting in a full cylindrical Al shell coating the wire \cite{Krogstrup2015}, as shown in Fig.~\ref{fig:1}(a). Subsequent fabrication used standard electron-beam lithography, deposition, etching, and liftoff, as described elsewhere \cite{Vaitiekenas2018}. Devices were operated in two configurations [Fig.~\ref{fig:1}(b)]: In the first configuration, four Au  contacts were made to the Al shell allowing four-wire resistance measurements; In the second, an additional tunneling contact to the InAs core at the end of the wire was used as a tunnel probe, giving local density of states, as discussed in Ref.~\cite{Vaitiekenas2018}. We investigated wires from three growth batches, denoted A, B, and C, with different core diameters, $d_{\rm C}$, and shell thicknesses, $t_{\rm S}$ (see Supplemental Material \cite{SupMaterial}). Transport measurements were carried out in a dilution refrigerator with a three-axis vector magnet and base temperature of 20~mK.

Carrier density in the InAs core is predominantly at the Al interface due to band bending \cite{Mikkelsen2018,Antipov2018}. Moreover, the density of carriers in Al is orders of magnitude higher than in InAs. We may therefore consider current to be carried by a hollow cylinder which is threaded by flux in an axial applied magnetic field. Induced circumferential supercurrents from the applied flux lead to Cooper pair breaking, characterized by the parameter $\alpha$, which controls the transition temperature $T_{\rm C}(\alpha)$, as described by Abrikosov-Gorkov expression,

\begin{figure}[t!]
\includegraphics[width=\linewidth]{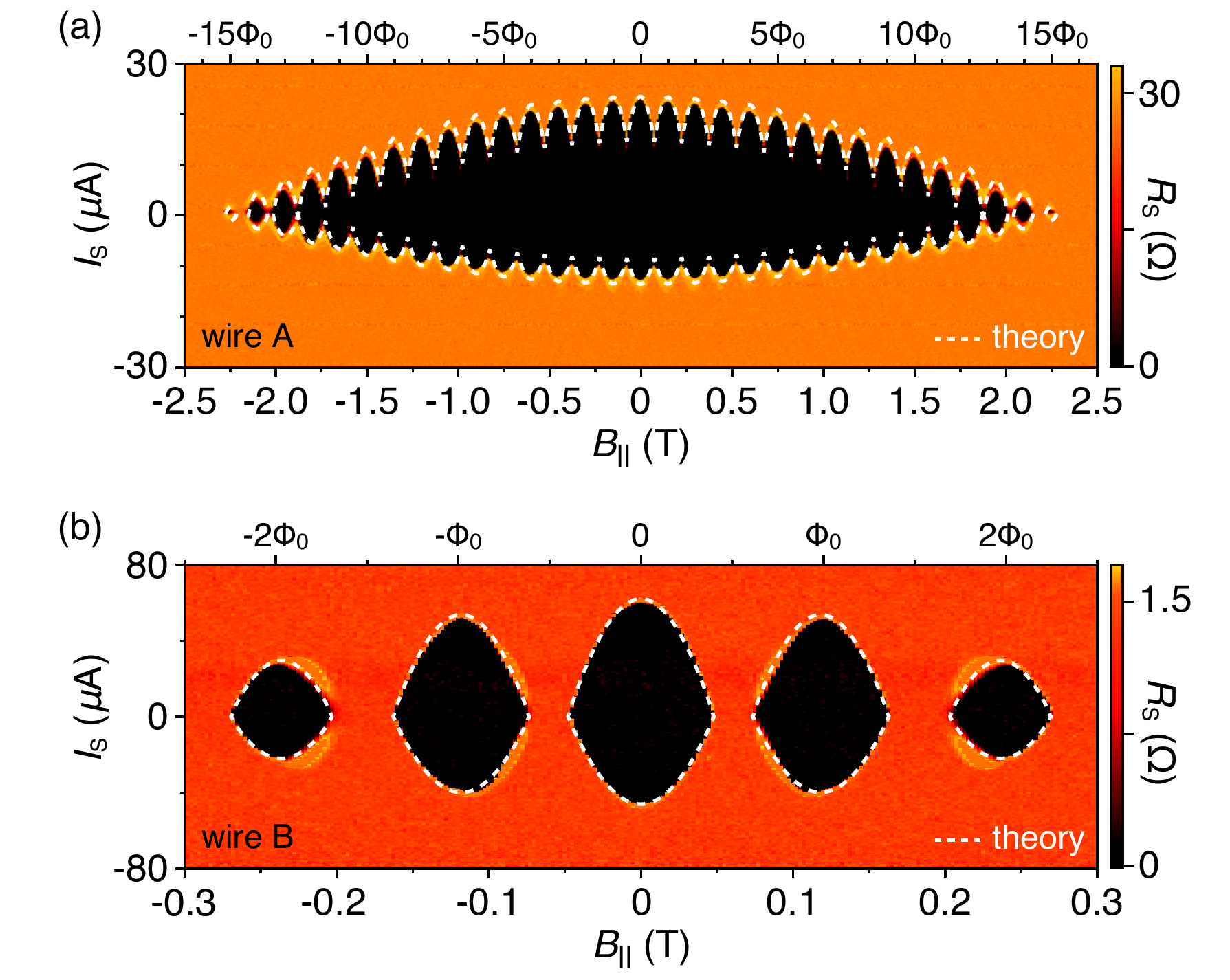}
\caption{\label{fig:3} (a) Shell resistance, $R_{\rm S}$, measured for wire A with shell thickness $t_{\rm S}=7$~nm as a function of axial magnetic field, $B_\parallel$, and current bias, $I_{\rm S}$. Both switching and re-trapping currents are periodically modulated by $B_\parallel$ up to $B_{\rm \parallel,C}=2.3$~T, whereafter the supercurrent is suppressed. The dashed theory curve is Eq.~\ref{eq:i_c} evaluated with $\alpha_\parallel$ from Eq. \ref{eq:alpha_paral} and the corresponding fit parameters measured for wire A. (b) Same as (a), but measured for wire B with shell thickness around $t_{\rm S}=25$~nm.}
\end{figure}

\begin{equation}\label{eq:digamma}
    \ln\left(\frac{T_{\rm C}(\alpha)}{T_{\rm C0}}\right) 
    = \Psi\left(\frac{1}{2}\right) 
    - \Psi\left( \frac{1}{2} + \frac{\alpha}{2\pi { }k_{\rm B} T_{\rm C}(\alpha)} \right),
\end{equation}

\noindent where $\Psi$ is the digamma function  \cite{Abrikosov1961}. Following  Refs.~\cite{Dao2009,Schwiete2009,Sternfeld2011}, the pair-breaking parameter for a hollow cylinder with wall thickness $t_{\rm S}$ in a parallel magnetic field $B_\parallel$ is given by

\begin{equation}\label{eq:alpha_paral}
    \alpha_{\parallel} = \frac{4\,\xi_{\rm S} { }^2 T_{\rm C0}}{A_{\rm F}} \left[\left( n-\frac{\Phi}{\Phi_0}
    \right)^2 + \frac{ t_{\rm S}{ }^2}{d_{\rm F}{ }^2} \left( \frac{\Phi{ }^2}{\Phi_0{ }^2}
    +\frac{n^2}{3} \right) \right],
\end{equation}

\noindent where $\xi_{\rm S}$ is the zero-field superconducting coherence length, $A_{\rm F}$ is the area of the cylinder cross section, the integer $n$ is the fluxoid quantum number, $\Phi$ is the applied flux, and $d_{\rm F}$ is the diameter of the cylinder [Fig.~\ref{fig:1}(a)]. Taking the dirty-limit expression for $\xi_{\rm S} ~=~\sqrt{\pi \hbar v_{\rm F} l_{\rm e}/24 k_{\rm B} T_{\rm C0}}$ with the Fermi velocity $v_{\rm F}$ and mean free path $l_{\rm e}$, we note that all parameters can either be measured directly from the micrograph of the device or from independent transport measurements (see Supplemental Material \cite{SupMaterial}).

Differential shell resistances, $R_{\rm S} = dV_{\rm S}/dI_{\rm S}$, for wires A and B are shown in Fig.~\ref{fig:2} as a function of $B_\parallel$ and temperature, $T$. Wires A and B have similar core diameters, $d_{\rm C} \sim 135$~nm, but different shell thicknesses. For wire A, with $t_{\rm S} = 7$~nm, $T_{\rm C}$ is finite throughout the measured range, and varies periodically with applied axial flux with amplitude $\sim 0.4$~K with no clear envelope reduction up to $B_\parallel = 0.4$~T. Normal-state resistance of the wire yields a coherence length $\xi_{\rm S} = 70$~nm, smaller than $d_{\rm C}$ (see Supplemental Material \cite{SupMaterial}). In contrast, wire B, with $t_{\rm S} = 25$~nm, has $\xi_{\rm S}  = 180$~nm $> d_{\rm C}$, and shows destructive regimes around $\pm \Phi_0/2$ and $\pm 3\Phi_0/2$. Resistances in these destructive regimes remain equal to the normal state resistance, $R_{\rm S} = R_{\rm N}$, to the lowest measured temperatures. 

The absence (presence) of the destructive regime in wire A (B) is consistent with the criterion of the superconducting coherence length being smaller (larger) than the wire diameter \cite{deGennes1981}. To be more quantitative, we plot in Fig.~\ref{fig:2} theoretical curves marking the superconductor-metal transition based on Eqs.~\ref{eq:digamma} and \ref{eq:alpha_paral} with independently measured wire parameters, using either the measured zero-field critical temperature, $T_{\rm C0}$ or, equivalently, the spectroscopically measured zero-field superconducting gap, $\Delta_0$, [Fig.~S1 in Supplemental Material \cite{SupMaterial}], which was consistent with the BCS relation $\Delta_{0} = 1.76\, k_{\rm B}T_{\rm C0}$ \cite{Tinkham1996}. Figure \ref{fig:2} demonstrates the remarkably good agreement found between experiment and theory. The observed increase of $T_{\rm C}$ with decreasing $t_{\rm S}$ is consistent with enhanced energy gaps for thin Al films \cite{Court2008}.

Similar to the effects of flux-induced circumferential supercurrent, a dc current, $I_{\rm S}$, applied along the wire can also drive the shell normal. The field-dependent critical current $I_{\rm C}(\alpha)$ can be related to the corresponding critical temperature, $T_{\rm C}(\alpha)$,

\begin{equation}\label{eq:i_c}
    I_{\rm C} (\alpha) = I_{\rm C0} \left(\frac{T_{\rm C}(\alpha)}{T_{\rm C0}} \right)^{3/2},
\end{equation}

\noindent where $I_{\rm C0}$ is the zero-field critical current \cite{Bardeen1962}.

\begin{figure}[t!]
\includegraphics[width=\linewidth]{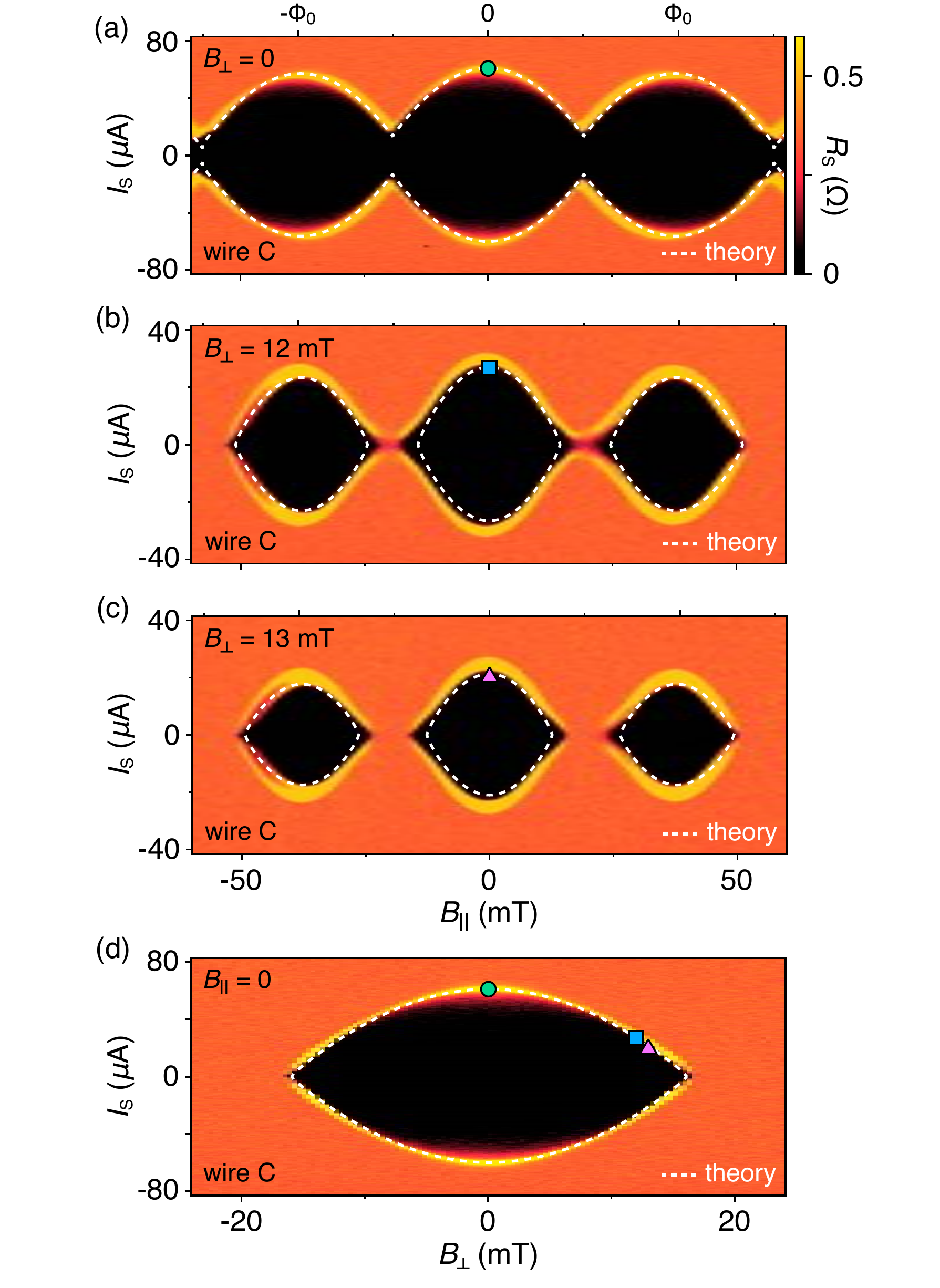}
\caption{\label{fig:4} (a) Base-temperature shell resistance, $R_{\rm S}$, measured for wire C as a function of current bias, $I_{\rm S}$, and parallel magnetic field, $B_\parallel$, at zero perpendicular magnetic field, $B_\perp = 0$. Approximately equal switching and re-trapping currents, which are proportional to the critical current, indicate nearly dissipationless supercurrent injection. The wire is non-destructive throughout the whole measured $B_\parallel$ range. (b) Same as (a), but at $B_\perp = 12$~mT. Around half-flux quantum and zero-current bias, an anomalous phase develops with a finite, but smaller than normal state resistance. (c) Same as (a), but measured at $B_\perp = 13$~mT. Around the half-flux quantum $R_{\rm S}$ remains at normal state value even at $I_{\rm S} = 0$. The theory curves in (a)-(c) are Eq.~\ref{eq:i_c} evaluated with $\alpha = \alpha_\parallel + \alpha_\perp$. (d) Critical current evolution as a function of $B_\perp$ measured at $B_\parallel = 0$. The theory curve in (d) is Eq.~\ref{eq:i_c} computed with $\alpha_\perp$.}
\end{figure}

\begin{figure*}[t!]
\includegraphics[width=\linewidth]{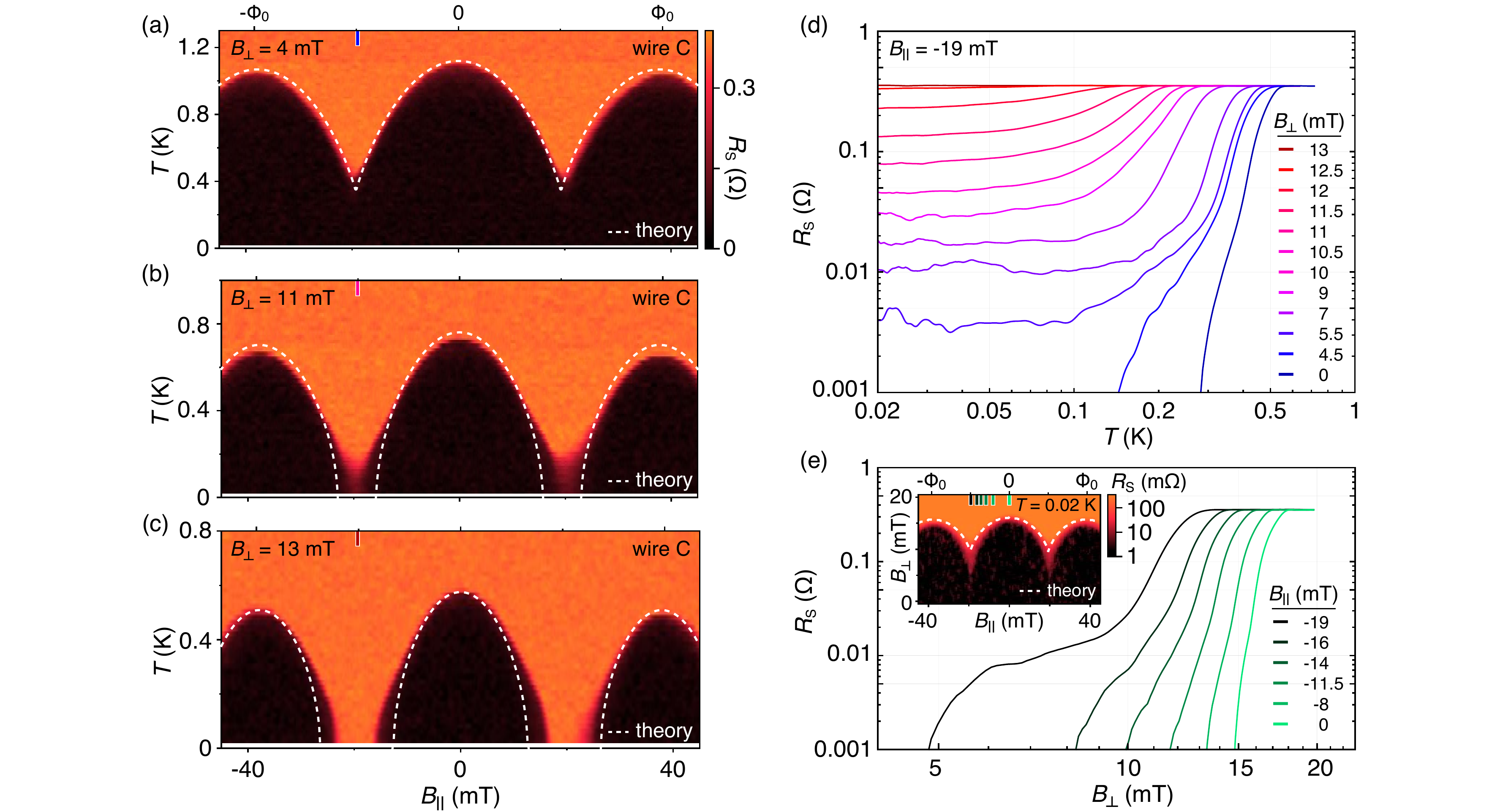}
\caption{\label{fig:5} (a) Differential shell resistance, $R_{\rm S}$, measured for wire C at $B_\perp = 4$~mT as a function of temperature, $T$, and parallel magnetic field, $B_\parallel$. For small $B_\perp$ the sample displays a non-destructive $T$--$B_\parallel$ phase diagram. (b) Same as (a), but measured at $B_\perp = 11$~mT. Around $\pm \Phi_0/2$ an anomalous-resistance phase develops at low $T$. (c) Same as (a), but measured at $B_\perp = 13$~mT. The shell resistance increases to the normal state value as the applied flux passes $\pm \Phi_0/2$, even at the base temperature. Note that $R_{\rm S}$ is finite for all temperatures above the mean-field theory predicted $T_{\rm C}$. The theory curves in (a)-(c) are Eq.~\ref{eq:digamma} numerically solved for $T_{\rm C} (\alpha_\parallel + \alpha_\perp)$. (d) Half-flux quantum $R_{\rm S}$ as a function of $T$ measured at different $B_\perp$ values. Close to $B_\perp = 0$, as the temperature is lowered, the sample displays a conventional normal-superconducting phase transition. Around $B_\perp = 5$~mT the shell resistance at low $T$ saturates to a finite, $B_\perp$-dependent value. Above $B_\perp = 13$~mT the shell resistance does not decrease below the normal state resistance. (e) Base-temperature $R_{\rm S}$ as a function of $B_\perp$ measured at different $B_\parallel$ or $\Phi$ values. The resistance increases with $B_\perp$ in a step-like manner with the step feature mostly pronounced at around $-\Phi_0/2$ of the applied flux. Inset: $R_{\rm S}$ as a function of $B_\perp$ and $B_\parallel$. The theory curve was computed using Eq.~\ref{eq:digamma}, where a critical $B_\perp$ was found for each $B_\parallel$, above which $T_{\rm C}$ vanishes.
}
\end{figure*}

Base-temperature $I_{\rm S}$--$B_\parallel$ phase diagrams for wires A and B are shown in Fig.~\ref{fig:3}. The data are taken sweeping from negative to positive bias, so show re-trapping currents for $I_{\rm S}<0$ and switching current for $I_{\rm S}>0$, both of which are proportional to the critical current, $I_{\rm C}$ \cite{Tinkham1996}. Similar to the transition temperature, $I_{\rm C}$ was observed to be $\Phi_0$-periodic in flux for both wires as expected from Eq.~\ref{eq:i_c}. For wire A, a bigger range of $B_\parallel$ [Fig.~\ref{fig:3}(a)] shows that the thin shell remains non-destructive up to $\sim 2$~T, corresponding to $\sim 13 \Phi_0$, then enters the destructive regime twice around $14 \Phi_0$ and finally turns fully normal around $B_{\rm \parallel,C} = 2.3$~T. 

Figure 3 shows theoretical curves based on Eqs.~1--3 superimposed on experimental data for both wire types. The zero-field switching and re-trapping currents were taken as input parameters, with other parameters measured independently. Again, excellent agreement between experiment and theory for both thin (wire A) and thick (wire B) shells was found.

We next consider the effects of an applied {\it transverse} magnetic field, $B_\perp$, which can be used to control a crossover between conventional and destructive Little-Parks regimes.  We investigate the combined effects of $B_{\parallel}$ and $B_\perp$ in wire C, with $d_{\rm C}~=~240$~nm and $t_{\rm S} = 40$~nm. The larger diameter reduces the field value $B_\parallel = \Phi_0/ A_{\rm F}$ and the thicker shell ensures a long $\xi_{\rm S} $, such that initially the wire is nearly destructive. The transition of the wire C from being non-destructive at $B_\perp = 0$ to destructive at $B_\perp = 13$~mT is depicted by $I_{\rm S}$--$B_\parallel$ phase diagrams in Fig.~\ref{fig:4}(a)-(c). 

Theoretically, the effect of $B_\perp$ on the superconducting transition  can be accounted for by introducing an additional pair-breaking term \cite{Shah2007},

\begin{equation}\label{eq:alpha_perp}
    \alpha_{\perp} = 
    \frac{4\, \xi_{\rm S} { }^2 T_{\rm C0}}{A_{\rm F}} \frac{\Phi_\perp{ }^2}{\Phi_0{ }^2},
\end{equation}

\noindent where $\Phi_\perp = B_\perp A_{\rm F}$. Figure~\ref{fig:4} shows the theoretical transitions based on Eqs.~1--4 using $\alpha = \alpha_\parallel + \alpha_\perp$ \cite{Rogachev2005} superimposed on experimental data.

Near the conventional-destructive crossover [Fig.~\ref{fig:4}(b)], a resistive state with $R_{\rm S}$ smaller than $R_{\rm N}$ was observed around $\pm\Phi_0/2$ and $I_{\rm S} = 0$. Figure~\ref{fig:5} examines this resistive state close to the crossover, around $B_{\perp}\sim 12$ mT, along with superimposed theory curves based on Eqs.~1--4. Note that unlike the situation far from the crossover [Fig.~\ref{fig:5}(a)], where theory and experiment agree well, in the vicinity of the crossover [Fig.~\ref{fig:5}(b,c)] mean-field theory predicted $T_{\rm C}$ deviates from the temperatures where the shell displays $R_{\rm N}$.

Temperature dependence of $R_{\rm S}$ around $-\Phi_0/2$ for several values $B_\perp$ near the conventional-destructive crossover are shown in Fig.~\ref{fig:5}(d). Throughout this regime, $R_{\rm S}$ was found to saturate to a temperature independent value, which can be tuned over two orders of magnitude with small changes in $B_\perp$. In contrast, a $R_{\rm S}$-$T$ trace taken close to the second destructive regime, not near a crossover ($B_\perp = 12$~mT and $B_\parallel = 52$~mT) remains temperature dependence down to the base temperature [Fig.~S2 in Supplemental Material \cite{SupMaterial}]. Qualitatively similar anomalous $R_{\rm S}$ saturation was also observed for different $B_\parallel$ values at a fixed $B_\perp$, see Fig.~S3 in Supplemental Material \cite{SupMaterial}. At base temperature the evolution of $R_{\rm S}$ as a function of $B_\perp$ shows a step-like increase, that is mostly pronounced around $\pm\Phi_0/2$, see Fig.~\ref{fig:5}(e).

A possible explanation for the saturation of  $R_{\rm S}$ in terms of disorder-induced variations of $\Delta$, separating the shell into normal and superconducting segments \cite{Dao2009} was tested by examining saturation effects in three segments of the same wire [Fig.~S4 in Supplemental Material \cite{SupMaterial}]. It was found that all segments behaved the same, arguing against long-range variation in $\Delta$ on the scale of the separation of contacts. We also note that the anomalous resistance develops predominantly above the theoretical $T_{\rm C}$, where the sample is expected to be in the normal state [Fig.~S5 in Supplemental Material \cite{SupMaterial}].

The step-like increase of $R_{\rm S}$ with $B_\perp$ shown in Fig.~\ref{fig:5}(e) is reminiscent of phase slips, similar to the ones reported in Refs.~\cite{Bezryadin2000,Rogachev2005}, except here they are activated by perpendicular field rather than temperature. This suggests a picture in which anomalous saturating resistance results from quantum fluctuations not captured by mean-field theory. In general, the probability of a transverse phase slip across a weak link is proportional to  $\exp{\left(-R_{\rm Q}/R_{\rm N}\right)}$, with the resistance quantum $R_{\rm Q}$, and therefore is exponentially small for wire C \cite{Vanevic2012}. However, near one half flux quantum, states with consecutive phase windings around the shell are degenerate, allowing quantum fluctuations to play a role. We note that both deep in the nondestructive regime [Fig.~\ref{fig:2}(a)] and deep in the destructive regime [Fig.~\ref{fig:2}(b)], no anomalous phase is observed.

Previous theoretical work \cite{Douglass1963,Dao2009} argued that the ratio of $d_{\rm F} t_{\rm S}/2$ to $\lambda^2$ controls the order of the superconductor-metal transition. The present experiments span the range, with wires A and B having $ d_{\rm F} t_{\rm S}/2 < \lambda^2$, whereas wire~C has $d_{\rm F} t_{\rm S}/2 \gtrsim \lambda^2$. We have not observed systematic qualitative difference across this ratio. A detailed investigation of the order of the transition, and its affect on the anomalous metallic phase, would make an interesting future study.

In summary, we have investigated destructive and nondestructive Little-Parks effect in InAs nanowires fully covered with epitaxial Al. Excellent agreement with Ginzburg-Landau mean-field theory was obtained across multiple reentrant quantum phase transitions using independently measured device and material parameters. Millitesla-scale perpendicular magnetic field was used to tune the crossover between destructive and non-destructive regimes, yielding an anomalous metallic phase around one-half flux quantum with a temperature independent resistance ranging over two orders of magnitude controlled by small changes in perpendicular field. This field-controllable anomalous phase is not explained by existing theory, but presumably involves quantum fluctuations between winding numbers of superconducting phase around the cylindrical superconducting shell.

We thank Mingtang Deng, Claus S\o rensen, and Shiv~Upadhyay for materials and experimental contributions, and Mikhail Feigelman, Karsten Flensberg, Steve Kivelson, Yuval Oreg, Gil Refael, and Boris Spivak for valuable discussions. Research was supported by Microsoft, the Danish National Research Foundation, and the European Commission.

\onecolumngrid
\clearpage
\onecolumngrid
\setcounter{figure}{0}
\setcounter{equation}{0}
\section{\large{S\MakeLowercase{upplemental} I\MakeLowercase{nformation}
}}
\renewcommand{\figurename}{FIG.~S}
\renewcommand{\tablename}{Table.~S}
\renewcommand{\thetable}{\arabic{table}}
\twocolumngrid

\section*{Methods}

\section*{Nanowire growth} The wires studied in this work were grown using molecular beam epitaxy on InAs(111)B substrate at $420~^\circ$C, via standard Au-catalized vapor-liquid-solid method. First, InAs wires were grown along the $[0001]$ direction with wurtzite crystal structure. Subsequently to the semiconductor growth, a full Al shell was grown at $-30~^\circ$C on all six facets by rotating the growth substrate with respect to the metal source, resulting in an epitaxial interface between the Al and InAs \cite{Krogstrup2015}. The core diameter was tuned by changing the Au seed particle size. The shell thickness was controlled by the Al growth time.

\section*{Device fabrication} For the device fabrication individual wires were transferred onto a degenerately n-doped Si substrate capped with a $200$~nm thermal oxide using a micro-manipulator station. Standard electron beam lithography techniques were used to pattern etching windows and contacts. A thin layer of AR 300-80 (new) adhesion promoter and double layer of EL6 copolymer resists was used to define the etching windows. The Al was then selectively wet-etch for $\sim 60$~s in MF-321 photoresist developer. To contact the Al shell, a stack of A4 and A6 PMMA resist was used. The Al oxide from under the contacts was removed by Ar-ion milling (RF ion source, $25$~W, $18$~mTorr, $9$~min) followed by normal Ti/Al ($5/210$~nm for wires A and B, and $5/350$~nm for wire C) ohmic contact metallization. To contact the InAs core a single layer of A6 PMMA resist was used. A gentler Ar-ion milling (RF ion source, $15$~W, $18$~mTorr, $6.5$~min) was used to remove the native oxide layer off the InAs core, followed by deposition of the normal Ti/Al ($5/180$~nm for wires A and B, and $5/350$~nm for wire C) ohmic contacts.

\begin{table}[b!]
\begin{tabular*}{\linewidth}{@{\extracolsep{\fill}}cccccc}
Wire & $d_{\rm F}$ (nm) & $d_{\rm M}$ (nm) & $d_{\rm C}$ (nm) & $t_{\rm S}$ (nm) & $L$ (nm)\\
\hline
A & 157$\pm$5 & 146$\pm$4 & 137$\pm$5 & 7$\pm$3 & 945$\pm$5\\
B & 195$\pm$5 & 163$\pm$4 & 135$\pm$5 & 24$\pm$3 & 945$\pm$5\\
C & 340$\pm$5 & 288$\pm$4 & 240$\pm$5 & 41$\pm$3 & 920$\pm$5\\
\end{tabular*}
\caption{\label{tb:wire_dim} Wire dimensions measured from micrographs. Full-wire diameter $d_{\rm F}$, mean diameter $d_{\rm M}$, core diameter $d_{\rm C}$, shell thickness $t_{\rm S}$, and distance between the voltage probes $L$.}
\end{table}

\begin{table}[b!]
\begin{tabular*}{\linewidth}{@{\extracolsep{\fill}}cccccc}
Wire & $R_{\rm N}$ ($\Omega$) & $T_{\rm C0}$ (K) & $\Delta_0~(\mu{\rm eV})$ & $I_{\rm S0}~(\mu A)$ & $I_{\rm R0}~(\mu A)$\\
\hline
A & 34.3$\pm$0.1 & 1.45$\pm$0.1 & 220$\pm$7 & 24$\pm$1 & 14$\pm$1\\
B & 1.6$\pm$0.1 & 1.22$\pm$0.1 & 183$\pm$3 & 62$\pm$2 & 46$\pm$2\\
C & 0.35$\pm$0.01 & 1.17$\pm$0.1 & 177$\pm$3 & 61$\pm$2 & 60$\pm$2\\
\end{tabular*}
\caption{\label{tb:wire_char} Wire characteristics extracted from transport measurements. Normal-state resistance $R_{\rm N}$, zero-field critical temperature $T_{\rm C0}$, superconducting gap $\Delta_0$, zero-field switching current $I_{\rm S0}$ and re-trapping current $I_{\rm R0}$.}
\end{table}

\begin{table}[b!]
\begin{tabular*}{\linewidth}{@{\extracolsep{\fill}}cccccc}
Wire & $\Delta B$ (mT) & $\rho$ ($\Omega$ nm) & $l_{\rm e}$ (nm) & $\xi_{\rm S} $ (nm) & $\lambda$ (nm)\\
\hline
A & 150$\pm$7 & 110$\pm$40 & 4$\pm$1 & 71$\pm$8 & 200$\pm$60\\
B & 120$\pm$5 & 20$\pm$3 & 20 $\pm$3 & 180$\pm$10 & 100$\pm$20\\
C & 38.4$\pm$0.9 & 14$\pm$1 & 29$\pm$2 & 224$\pm$8 & 89$\pm$7\\
\end{tabular*}
\caption{\label{tb:wire_quant} Calculated wire quantities. Flux period in magnetic field $\Delta B = \Phi_0/A_{\rm M}$, resistivity $\rho$, mean free path $l_{\rm e}$, zero-field superconducting coherence length $\xi_{\rm S}$ and Ginzburg-Landau penetration depth $\lambda$.}
\end{table}

\section*{Wire parameters} The main wire parameters, including the ones use to compute the theory curves in the main text, are summarized in Tables~S\ref{tb:wire_dim}, S\ref{tb:wire_char} and S\ref{tb:wire_quant}. The full-wire diameter, $d_{\rm F}$, and the core diameter, $d_{\rm C}$, [Fig.~1(a) in the main-text] as well as the distance between the voltage probes, $L$, for each wire were measured from individual micrographs. For all the wires the Al oxide was assumed to be $t_{\rm Ox} = 2$~nm. Using simple trigonometrical considerations one can deduce the full cross-sectional area $A_{\rm F} = 3 \sqrt{3} (d_{\rm F} - 2 t_{\rm Ox})^2/8$, the shell thickness $t_{\rm S}=\sqrt{3}\left(d_{\rm F}-d_{\rm C}\right)/4 - t_{\rm Ox}$ and the mean wire diameter $d_{\rm M} = (d_{\rm F} - 2t_{\rm Ox} + d_{\rm C})/2$. The normal state resistance $R_{\rm N}$ and the zero-field transition temperature $T_{\rm C0}$ were measured while cooling down the sample. Zero-field switching $I_{\rm S0}$ and re-trapping $I_{\rm R0}$ currents were measured at the base temperature. The period of the Little-Park oscillations in magnetic field can be calculated using $\Delta B = \Phi_0/A_{\rm M} = 8\, \Phi_0/ 3\sqrt{3}\, d_{\rm M}^2$. The shell resistivity is given by $\rho= R_{\rm N} (A_{\rm F} - A_{\rm C})/L$, where $A_{\rm C} = 3 \sqrt{3} d_{\rm C}^2/2$ is the core cross-sectional area. The Drude mean free path for electrons in the shell is determined using $l_{\rm e} = m_{\rm e}{ }v_{\rm F}/ e^2 n \rho$, with electron mass $m_{\rm e}$, electron Fermi velocity in Al $v_{\rm F} = 2.03 \times 10^6$~m/s \cite{Kittel2005}, electron charge $e$ and charge carrier density $n = k_{\rm F}^3/ 3 \pi^2$, where $k_{\rm F}$ is the Fermi wave vector. The dirty-limit superconducting coherence length is given by \cite{Tinkham1996} $\xi_{\rm S} ~=~\sqrt{\pi \hbar v_{\rm F}  l_{\rm e}/24 k_{\rm B} T_{\rm C0}}$, where $\hbar$ is the reduced Planck constant and $k_{\rm B}$ is the Boltzmann constant. For a dirty superconductor, the Ginzburg-Landau penetration depth is \cite{Tinkham1996} $\lambda(T) = \lambda_{\rm L}(T) \sqrt{\xi_0/1.33{ }l_{\rm e}}$, with the London penetration depth $\lambda_{\rm L} (T) = \lambda_{\rm L} (0) / \sqrt{2(1-T/T_{\rm C0})}$, and the coherence length is $\xi_{\rm S} (T) = 0.855 \sqrt{\xi_0 l_{\rm e}/(1-T/T_{\rm C0})}$. This gives $\lambda = \lambda_{\rm L} \xi_{\rm S}/1.39{ }l_{\rm e}$, with the zero-temperature London penetration depth $\lambda_{\rm L} = 16$~nm \cite{Kittel2005}.

\begin{figure}[t!]
\includegraphics[width=\linewidth]{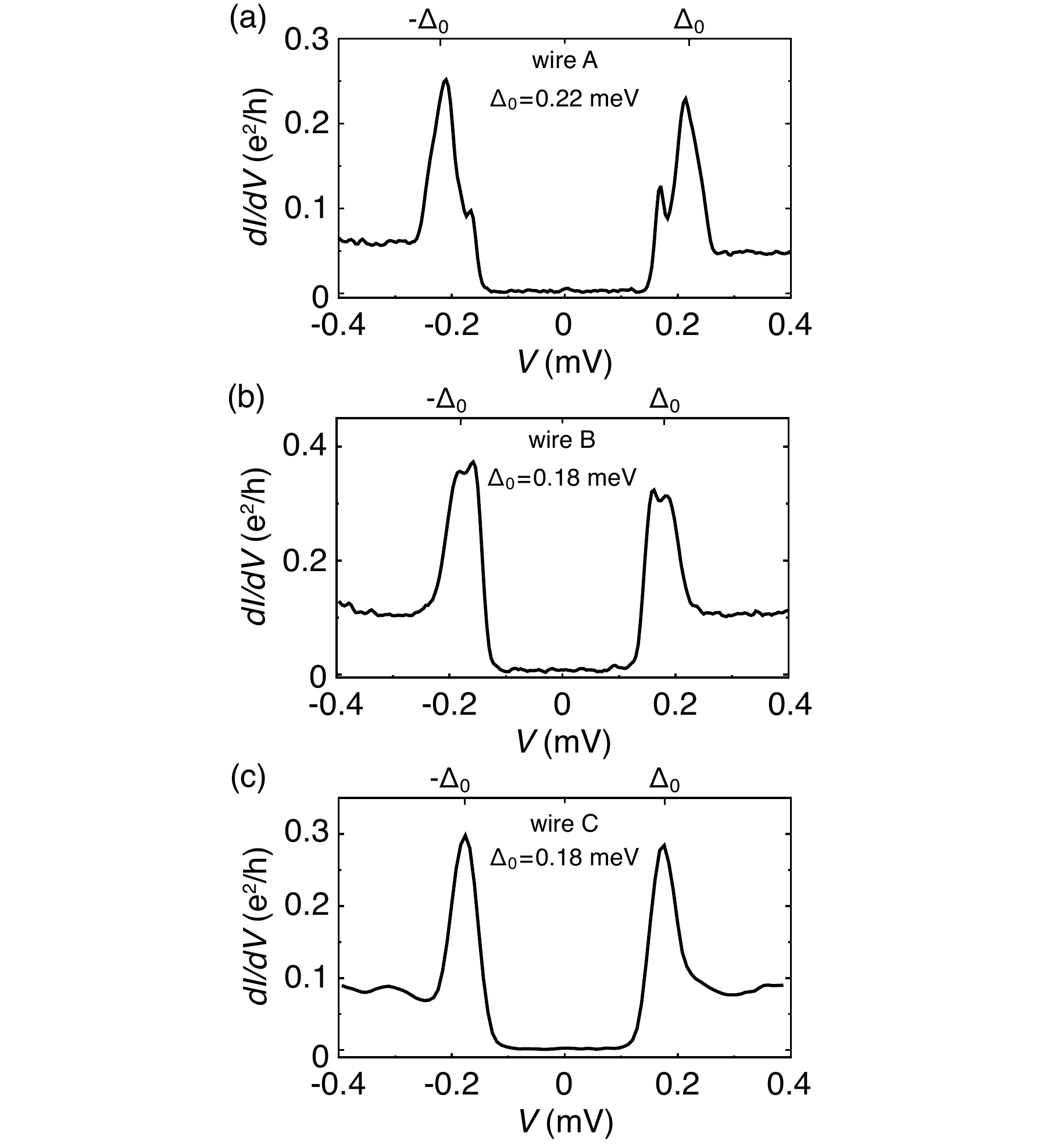}
\caption{\label{fig:S1} Differential conductance, $dI/dV$, as a function of source-drain voltage bias, $V$, measured for (a) wire A, (b) wire B, and (c) wire C.}
\end{figure}

\section*{Density of states} Each of the measured wire is equipped with a tunneling probe at its end, see the main-text Fig.~1(b). Applying voltage to the back-gate, $V_{\rm BG}$, creates a tunnel barrier in the bare-semiconducting (InAs) segment, seperating the normal-metal (Ti/Au) contact and the wire (Al/InAs). In the tunneling regime, the change in the current through the junction with the applied voltage bias corresponds to the local density of states. Differential tunnelling conductance, $dI/dV$, measured for all three wires as a function of source-drain voltage, $V$, is shown in Fig.~S1. For wire A, with the thinnest shell, the measured superconducting gap is $\Delta_0 = 220~\mu$eV, whereas both wires B and C show a gap of around $\Delta_0 = 180~\mu$eV. All three gaps agree (within the experimental error) with the BCS theory predicted value $\Delta_0 = 1.764{ }k_{\rm B}{ }T_{\rm C0}$. Wires A and B display additional peaks in density of states at the energies below the main superconducting gap. We identify these with the proximity induced gaps inside the semiconducting cores.

\begin{figure}[t!]
\includegraphics[width=\linewidth]{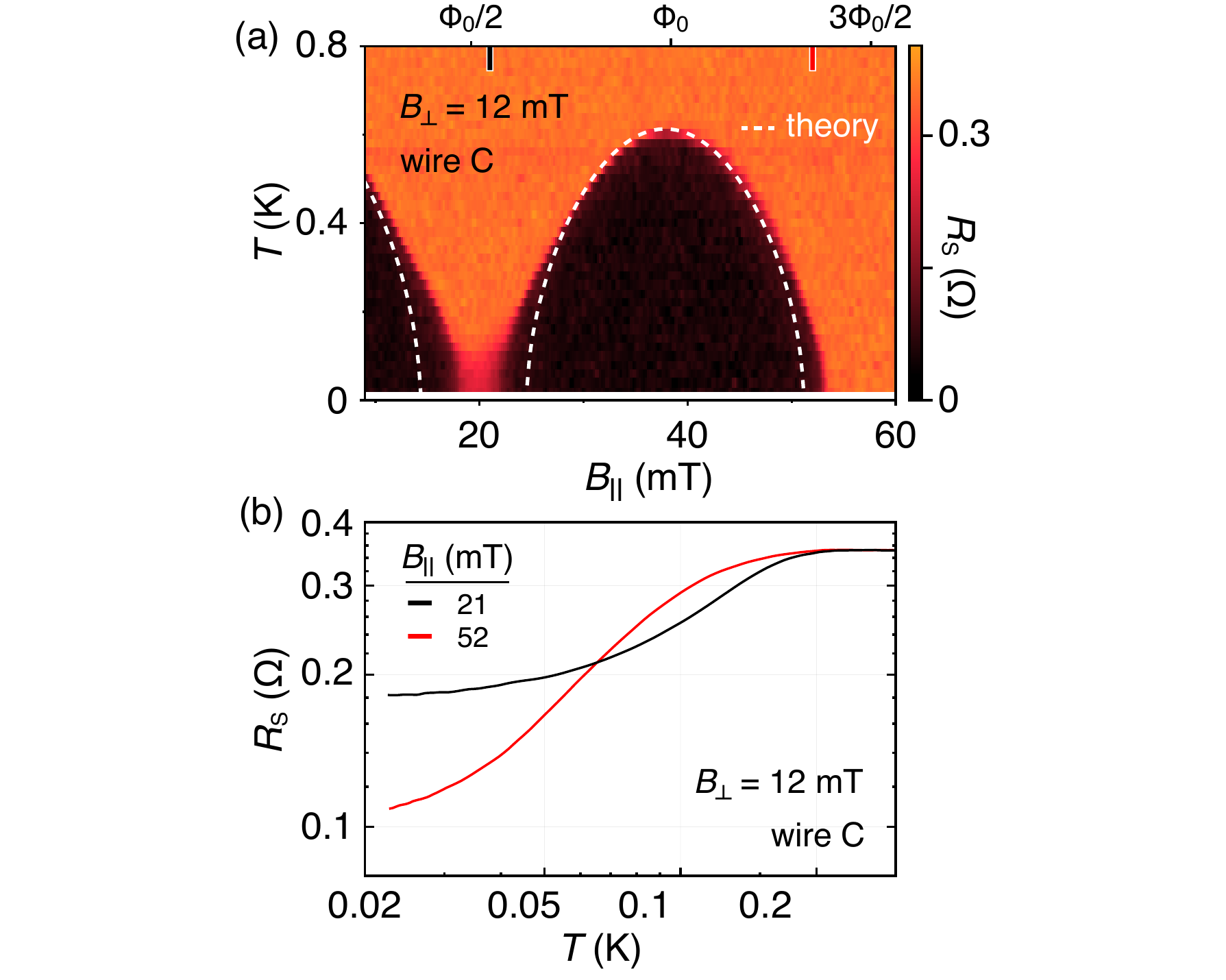}
\caption{\label{fig:S2} (a) Differential shell resistance, $R_{\rm S}$, as a function of parallel magnetic field, $B_\parallel$, and temperature, $T$, measured for wire C at perpendicular magnetic field $B_\perp = 12$~mT. The theory curve is the main-text Eq.~3 computed with $\alpha = \alpha_\parallel + \alpha_\perp$. (b) $R_{\rm S}$-$T$ traces measured at $B_\perp = 12$~mT. At $B_\parallel = 21$~mT, close to $\Phi_0/2$ applied flux, $R_{\rm S}$ saturates at low temperatures. At $B_\perp = 51$~mT, before the wire enter the destructive regime around $3\Phi_0/2$, $R_{\rm S}$ shows temperature dependence even at the base temperature.}
\end{figure}

\begin{figure}[t!]
\includegraphics[width=\linewidth]{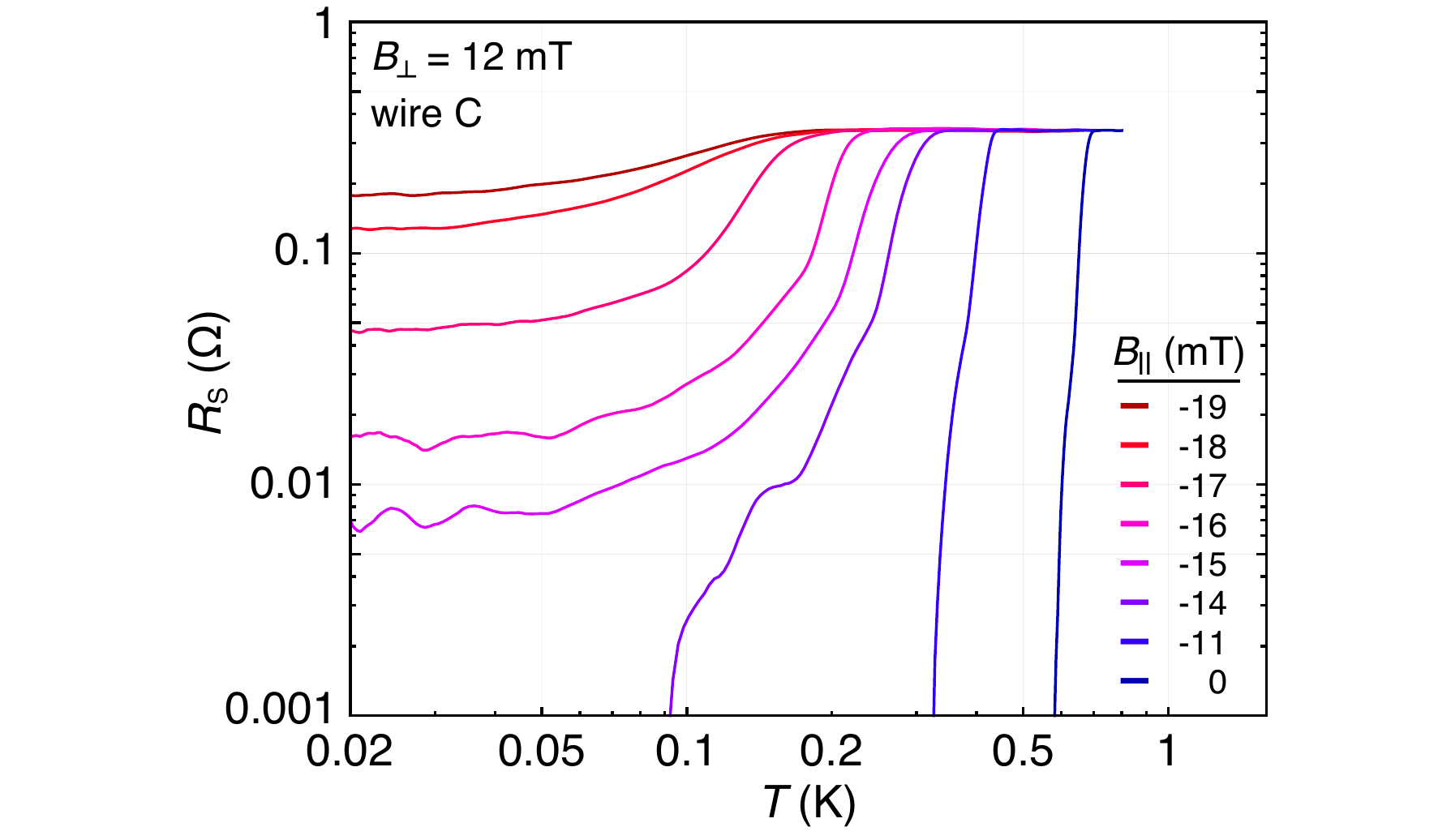}
\caption{\label{fig:S3} Differential shell resistance, $R_{\rm S}$ as a function of $T$ measured at fixed $B_\perp = 12$~mT for wire C at different $B_\parallel$ values. Around $B_\parallel = 0$, as $T$ is lowered, the sample displays a conventional normal-superconducting phase transition. As the field is tuned to $B_\parallel = -14$~mT the shell resistance starts to saturate at low $T$ to a finite, $B_\parallel$-dependent value.}
\end{figure}

\begin{figure}[t!]
\includegraphics[width=\linewidth]{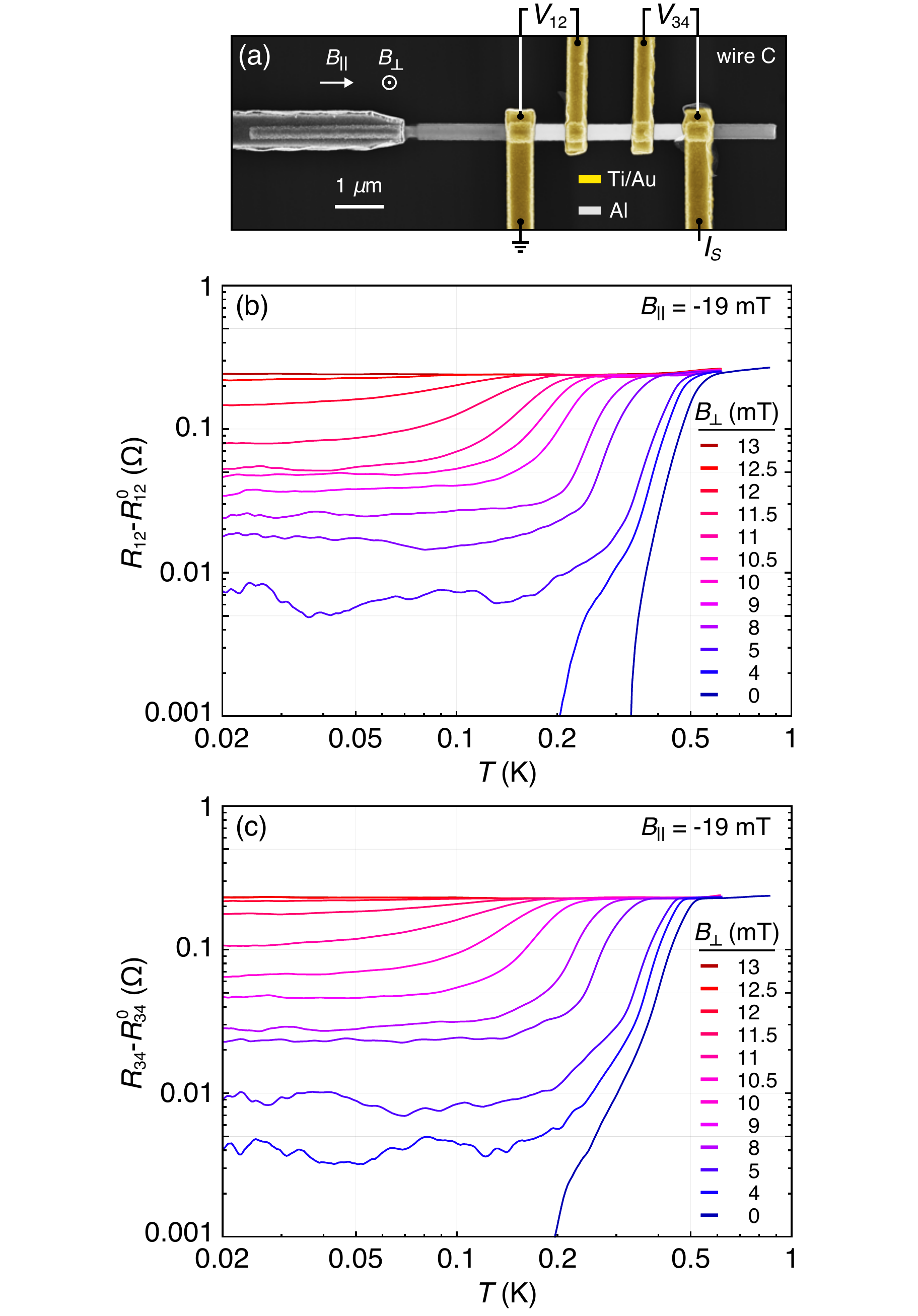}
\caption{\label{fig:S4} (a) Micrograph of wire C with the highlighted three-terminal setups for the outer segments shell resistance measurements. (b) Differential shell resistance in the left-most wire C segment, $R_{12} = dV_{12}/dI_{\rm S}$ (with the subtracted contact resistance $R_{12}^0$) measured as a function of temperature $T$ at $B_\parallel = -19$~mT and different $B_\perp$ values. (c) Similar to (b), but measured for the right-most segment. The contact resistances $R_{12}^0$ and $R_{34}^0$ were measured around the base temperature at $B_\perp = 0$ and $B_\parallel = -19$~mT.}
\end{figure}

\begin{figure}[t!]
\includegraphics[width=\linewidth]{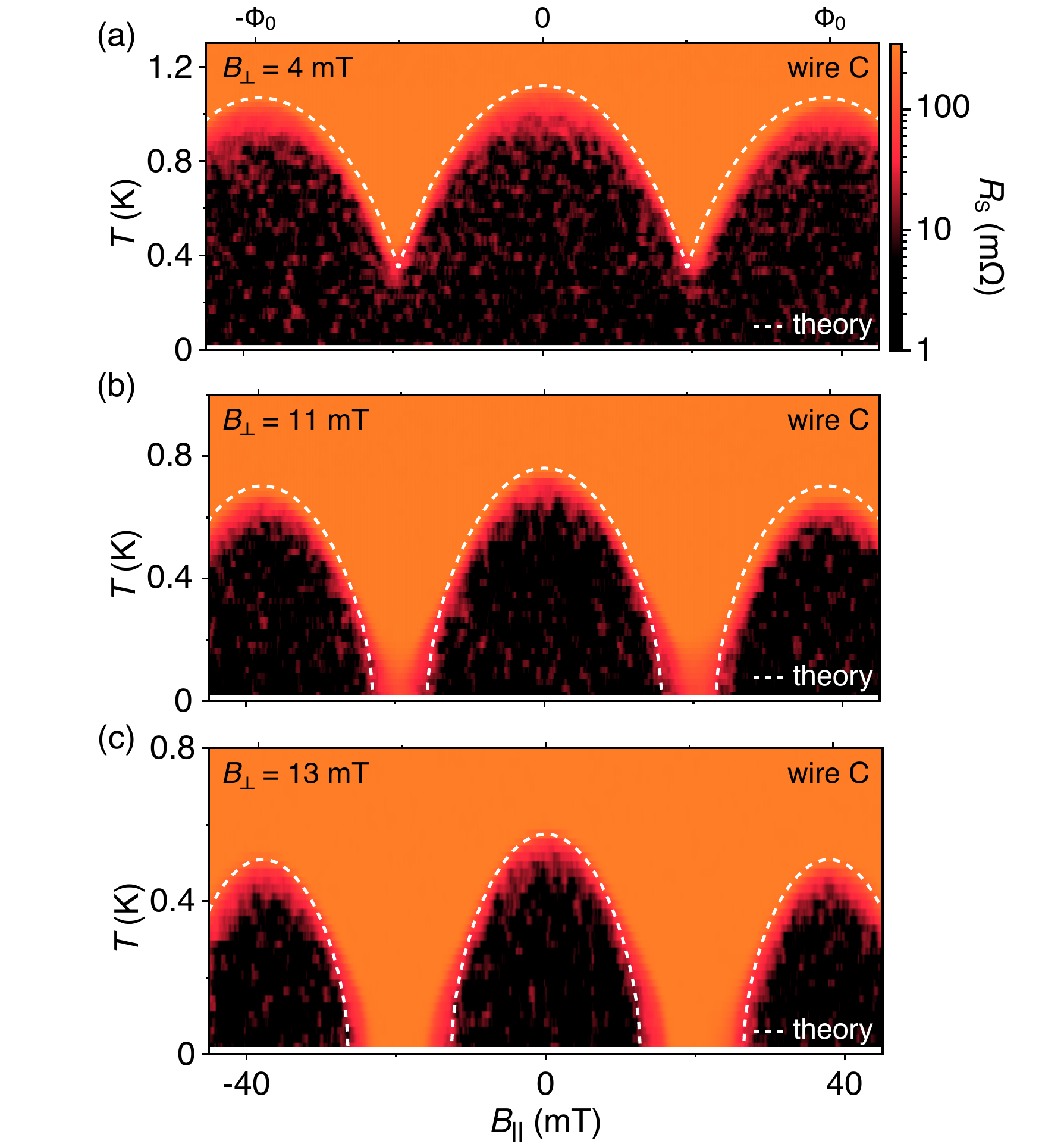}
\caption{\label{fig:S5} Same data as in the main-text Fig.~5(a), (b) and (c), but in logarithmic color scale highlighting the low-resistance features. At low $T$, the anomalous phase is present predominantly above the $T_{\rm C}$ predicted by the mean-field theory. The finite resistance at higher temperatures, below the arcs of the theory curves, arise presumably due to thermal fluctuations.}
\end{figure}

\section*{Non-saturating resistance} The observed low-temperature saturation of the half-flux quantum $R_{\rm S}$ might rise a question whether it is not an artifact of a deficient cooling. In other words, if the electron temperature upon cooling would saturate at some elivated temperature, so would the shell resistance. To rule out such an explanation we record two $R_{\rm S}$-$T$ traces for wire C at $B_\perp = 12$~mT, see Fig.~S2: One at $B_\parallel = 21$~mT, close to $\Phi_0/2$ applied flux quantum, displaying the anomalous $R_{\rm S}$ saturation; Another at $B_{\parallel} = 52$~mT, before the destructive regime around $3\Phi_0/2$, with a $T$-dependent $R_{\rm S}$ down to the base temperature. Furthermore, the data in the main-text Fig.~5(d) shows that the $R_{\rm S}$ starts to saturate at different temperatures for different $B_\perp$. Finally, it is unlikely for a poor electron cooling to cause the observed broadening of the anomalous phase in flux, see the main-text Fig.~5(b) and (c), as well as Fig. S2(a).


\section*{Flux-dependent resistance saturation} The data shown in the main-text Fig.~5 demonstrate that at a fixed $B_\parallel = -19$~mT (around $-\Phi_0/2$ of the applied flux) $R_{\rm S}$ at low $T$ saturates to a $B_\perp$-dependent value. We observe a qualitatively similar $B_\parallel$-dependent anomalous saturation of $R_{\rm S}$ at a fixed $B_\perp = 12$~mT, see Fig.~S3.


\section*{Outer segments} To demonstrate that the anomalous resistance saturation shown in the main-text Fig.~5(d) is not due to a local disorder in the middle-wire segment, we investigate the outer two wire segments using three-terminal setup, see Fig.~S4(a). Differential shell resistances $R_{12} = dV_{12}/dI_{\rm S}$ and $R_{34} = dV_{34}/dI_{\rm S}$ with the subtracted corresponding contact resistances measured as a function of $T$ at $B_\parallel = -19$~mT and different $B_\perp$ values are shown in Fig.~S4(b) and (c). The contacts resistances $R_{12}^0$ and $R_{34}^0$ were measured around the base temperature at $B_\perp = 0$ and $B_\parallel = -19$~mT. The observed $B_\perp$-dependent, low-temperature anomalous shell resistances are qualitatively similar to the $R_{\rm S}$ of the middle segment. The small quantitative discrepancies between the segments might arise due to the uncertainty in the applied $B_\perp$ or a small wire tapering.

\section*{Anomalous phase vs. mean-field theory} Figure S5 shows the same data as in the main-text Fig.~5(a)-(c), but plotted in a logarithmic color scale to highlight the low-resistance features. It appears that the anomalous resistance phase at low $T$ develops predominantly above the mean-field theory predicted $T_{\rm C}$, close to the $\pm \Phi_0/2$ of the applied flux. At elevated $T$, the wire shows finite, but reduced $R_{\rm S}$ around $0$ and $\pm \Phi_0$ of the applied flux, presumably arising due to thermal fluctuations.


\begin{thebibliography}{99}

\bibitem{Sondhi1997} S.~L.~Sondhi, S.~M.~Girvin, J.~P.~Carini, and D.~Shahar, Rev.~Mod.~Phys. \textbf{69}, 315 (1997).

\bibitem{Vojta2000} T.~Vojta, Ann.~Phys. \textbf{9}, 403 (2000).

\bibitem{Shah2007} N.~Shah and A.~Lopatin, Microscopic, Phys.~Rev.~B \textbf{76}, 094511 (2007).

\bibitem{Si2010} Q.~Si and F.~Steglich, Science \textbf{329}, 1161 (2010).

\bibitem{Norman2011} M.~R.~Norman, Science \textbf{332}, 196 (2011).

\bibitem{Tinkham1996} M.~Tinkham, \textit{Introduction to Superconductivity} (McGraw Hill, New York, ed. 2, 1996).

\bibitem{DelMaestro2009} A.~Del~Maestro, B.~Rosenow and S.~Sachdev, Ann.~Phys. \textbf{324}, 523 (2009).

\bibitem{Goldman2010} A.~M.~Goldman, Int. J. Mod. Phys. B \textbf{24}, 4081 (2010).

\bibitem{Kapitulnik2019} A.~Kapitulnik, S.~A.~Kivelson, and B.~Spivak, Rev.~Mod.~Phys. \textbf{91}, 011002 (2019).

\bibitem{Bezryadin2000} A.~Bezryadin, C.~N.~Lau, and M.~Tinkham, Nature \textbf{404}, 971 (2000).

\bibitem{Zaikin1997} A.~D.~Zaikin, D.~S.~Golubev, A.~van~Otterlo, and G.~T.~Zim\'{a}nyi, Phys.~Rev.~Lett. \textbf{78}, 1552 (1997).

\bibitem{Astafiev2012} O.~V.~Astafiev, L.~B.~Ioffe, S.~Kafanov, Yu.~A.~Pashkin, K.~Yu.~Arutyunov, D.~Shahar, O.~Cohen, and J.~S.~Tsai, Nature \textbf{484}, 355 (2012).

\bibitem{Mooij2005} J.~E.~Mooij and C.~J.~P.~M.~Harmans, N.~J.~Phys. \textbf{7}, 219 (2005).

\bibitem{Deaver1961} B.~S.~Deaver~Jr. and W.~M.~Fairbanks, Phys.~Rev.~Lett. \textbf{7}, 43 (1961).

\bibitem{Doll1961} H.~Doll and M.~N{\"a}bauer, Phys.~Rev.~Lett. \textbf{7}, 51 (1961).

\bibitem{Douglass1963} D.~H.~Douglass~Jr. Phys.~Rev. \textbf{132}, 513 (1963).

\bibitem{Little1962} W.~A.~Little and R.~D.~Parks, Phys.~Rev.~Lett. \textbf{9}, 9 (1962).

\bibitem{deGennes1981} P.-G.~de~Gennes,  C.~R.~Acad.~Sci. \textbf{292}, 279 (1981).

\bibitem{Arutyunyan1980} R.~M.~Arutyunyan and G.~F.~Zharkov, Zh. Eksp. Teor. Fiz. \textbf{79}, 245 (1980).

\bibitem{Schwiete2009} G.~Schwiete and Y.~Oreg, Phys.~Rev.~Lett. \textbf{103}, 037001 (2009).

\bibitem{Liu2001} Y.~Liu, Yu.~Zadorozhny, M.~M.~Rosario, B.~Y.~Rock, P.~T.~Carrigan, and H.~Wang, Sience \textbf{294}, 2332 (2001).

\bibitem{Sternfeld2011} I.~Sternfeld, E.~Levy, M.~Eshkol, A.~Tsukernik, M.~Karpovski, H.~Shtrikman, A.~Kretinin, and A.~Palevski, Phys.~Rev.~Lett. \textbf{107}, 037001 (2011).

\bibitem{Dao2009} V.~H.~Dao and L.~F.~Chibotaru, Phys.~Rev.~B \textbf{79}, 134524 (2009).

\bibitem{Vafek2005} O.~Vafek, M.~R.~Beasley, and S.~A. ~Kivelson, https://arxiv.org/abs/cond-mat/0505688 (2005).

\bibitem{Lopatin2005} A.~V.~Lopatin, N.~Shah, and V.~M.~Vinokur, Phys.~Rev.~Lett. \textbf{94}, 037003 (2005).

\bibitem{Krogstrup2015} P.~Krogstrup, N.~L.~B.~Ziino, W.~Chang, S.~M.~Albrecht, M.~H.~Madsen, E.~Johnson, J.~Nyg\aa rd, C.~M.~Marcus, and T.~S.~Jespersen, Nat.~Mater. \textbf{14}, 400 (2015).

\bibitem{Vaitiekenas2018} S.~Vaitiek\.enas, M.-T.~Deng, P.~Krogstrup, and C.~M.~Marcus, https://arxiv.org/abs/1809.05513 (2018).

\bibitem{SupMaterial} See Supplemental Material for more detailed device description and additional measurements.

\bibitem{Kittel2005} C.~Kittel, \textit{Introduction to Solid State Physics} (Wiley, ed. 8, 2005).

\bibitem{Mikkelsen2018} E.~G.~Mikkelsen, P.~Kotetes, P.~Krogstrup, and K.~Flensberg, Phys.~Rev.~X \textbf{8}, 031040 (2018).

\bibitem{Antipov2018} A.~E.~Antipov, A.~Bargerbos, G.~W.~Winkler, B.~Bauer, E.~Rossi, and R.~M.~Lutchyn, Phys.~Rev.~X \textbf{8}, 031041 (2018).

\bibitem{Abrikosov1961} A.~A.~Abrikosov and L.~P.~Gorkov, Sov.~Phys.~JETP \textbf{12}, 1243 (1961).

\bibitem{Court2008} N.~A.~Court, A.~J.~Ferguson and R.~G.~Clark, Supercond.~Sci.~Technol. \textbf{21}, 015013 (2008).

\bibitem{Bardeen1962} J.~Bardeen, Rev.~Mod.~Phys. \textbf{34}, 667 (1962).

\bibitem{Rogachev2005} A.~Rogachev, A.~T.~Bollinger, and A.~Bezryadin, Phys.~Rev.~Lett. \textbf{94}, 017004 (2005).

\bibitem{Vanevic2012} M.~Vanevi\'{c} and Y.~V.~Nazarov, Phys.~Rev.~Lett. \textbf{108}, 187002 (2012).

\end{thebibliography}

\begin{thebibliography}{99}

\makeatletter
\makeatother

\bibitem{Krogstrup2015} P.~Krogstrup, N.~L.~B.~Ziino, W.~Chang, S.~M.~Albrecht, M.~H.~Madsen, E.~Johnson, J.~Nyg\aa rd, C.~M.~Marcus, and T.~S.~Jespersen, Nat.~Mater. \textbf{14}, 400 (2015).

\bibitem{Kittel2005} C.~Kittel, \textit{Introduction to Solid State Physics} (Wiley, ed. 8, 2005).

\bibitem{Tinkham1996} M.~Tinkham, \textit{Introduction to Superconductivity} (McGraw Hill, New York, ed. 2, 1996).

\end{thebibliography}
\end{document}